\documentclass[conference]{IEEEtran}

\newcommand{\pdftitle}{
Multi-agent Reinforcement Learning-based In-place Scaling Engine for Edge-cloud Systems}

\usepackage[nolist,nohyperlinks]{acronym}
\usepackage{amsmath}
\usepackage{amsfonts}
\usepackage{breqn}
\usepackage{balance}
\usepackage{cite}
\usepackage{framed}
\usepackage{siunitx}
\usepackage{soul}
\usepackage{multirow}
\usepackage{etoolbox}
\usepackage{algorithm}
\usepackage{algorithmic}

\usepackage{textcomp}

\usepackage{tikz}

\usepackage{makecell}

\definecolor{JovanGreen}{RGB}{34,139,34}
\newcommand{\green}[1]{\textcolor{JovanGreen}{#1}}
\newcommand{\red}[1]{\textcolor{red}{#1}}

\usepackage{booktabs}

\usepackage{blindtext}

\usepackage{graphicx}
\usepackage{subcaption}
\usepackage{color}

\usepackage{amssymb}
\usepackage{pifont}

\graphicspath{{./images/}}

\usepackage{url}
\usepackage[bookmarks=false]{hyperref}
\hypersetup{
  backref,
  hidelinks,
  colorlinks=false,
  breaklinks=true,
  bookmarksopen=false,
  pdftitle=\pdftitle,
  pdfauthor={Author}
}

\newtoggle{authorcomment}
\toggletrue{authorcomment}

\newcommand{\cmark}{\ding{51}}%
\newcommand{\xmark}{\ding{55}}%

\newcommand\copyrighttext{%
  \footnotesize \textcopyright 2025 IEEE. Personal use of this material is permitted.
  Permission from IEEE must be obtained for all other uses, in any current or future
  media, including reprinting/republishing this material for advertising or promotional
  purposes, creating new collective works, for resale or redistribution to servers or
  lists, or reuse of any copyrighted component of this work in other works.
  }
\newcommand\copyrightnotice{%
\begin{tikzpicture}[remember picture,overlay]
\node[anchor=north,yshift=-10pt] at (current page.north) {\fbox{\parbox{\dimexpr\textwidth-\fboxsep-\fboxrule\relax}{\copyrighttext}}};
\end{tikzpicture}%
}

\begin{document}
\bstctlcite{IEEEexample:BSTcontrol}

\title{\pdftitle}


\author{\IEEEauthorblockN{Jovan Prodanov\IEEEauthorrefmark{1}, Bla\v{z} Bertalani\v{c}\IEEEauthorrefmark{1},  Carolina Fortuna\IEEEauthorrefmark{1}, Shih-Kai Chou\IEEEauthorrefmark{1}, Matja\v{z} Branko Juri\v{c}\IEEEauthorrefmark{2}, \\ Ramon Sanchez-Iborra\IEEEauthorrefmark{3} and Jernej Hribar\IEEEauthorrefmark{1}} \IEEEauthorblockA{\IEEEauthorrefmark{1}Jožef Stefan Institute, Ljubljana, Slovenia. \\ 
\IEEEauthorrefmark{2}Faculty of Computer and Information Science, University of Ljubljana, Ljubljana, Slovenia\\
\IEEEauthorrefmark{3}Department of Information and Communication Engineering, University of Murcia, Murcia, Spain  
\\ Email: \{blaz.bertalanic, carolina.fortuna, shih-kai.chou, jernej.hribar\}@ijs.si 
}
}

\maketitle

\begin{acronym}[MACHU]
  \acro{iot}[IoT]{Internet of Things}
  \acro{cr}[CR]{Cognitive Radio}
  \acro{ofdm}[OFDM]{orthogonal frequency-division multiplexing}
  \acro{ofdma}[OFDMA]{orthogonal frequency-division multiple access}
  \acro{scfdma}[SC-FDMA]{single carrier frequency division multiple access}
  \acro{rbi}[RBI]{ Research Brazil Ireland}
  \acro{rfic}[RFIC]{radio frequency integrated circuit}
  \acro{sdr}[SDR]{Software Defined Radio}
  \acro{sdn}[SDN]{Software Defined Networking}
  \acro{su}[SU]{Secondary User}
  \acro{ra}[RA]{Resource Allocation}
  \acro{qos}[QoS]{quality of service}
  \acro{usrp}[USRP]{Universal Software Radio Peripheral}
  \acro{mno}[MNO]{Mobile Network Operator}
  \acro{mnos}[MNOs]{Mobile Network Operators}
  \acro{gsm}[GSM]{Global System for Mobile communications}
  \acro{tdma}[TDMA]{Time-Division Multiple Access}
  \acro{fdma}[FDMA]{Frequency-Division Multiple Access}
  \acro{gprs}[GPRS]{General Packet Radio Service}
  \acro{msc}[MSC]{Mobile Switching Centre}
  \acro{bsc}[BSC]{Base Station Controller}
  \acro{umts}[UMTS]{universal mobile telecommunications system}
  \acro{Wcdma}[WCDMA]{Wide-band code division multiple access}
  \acro{wcdma}[WCDMA]{wide-band code division multiple access}
  \acro{cdma}[CDMA]{code division multiple access}
  \acro{lte}[LTE]{Long Term Evolution}
  \acro{papr}[PAPR]{peak-to-average power rating}
  \acro{hn}[HetNet]{heterogeneous networks}
  \acro{phy}[PHY]{physical layer}
  \acro{mac}[MAC]{medium access control}
  \acro{amc}[AMC]{adaptive modulation and coding}
  \acro{mimo}[MIMO]{multiple input multiple output}
  \acro{rats}[RATs]{radio access technologies}
  \acro{vni}[VNI]{visual networking index}
  \acro{rbs}[RB]{resource blocks}
  \acro{rb}[RB]{resource block}
  \acro{ue}[UE]{user equipment}
  \acro{cqi}[CQI]{Channel Quality Indicator}
  \acro{hd}[HD]{half-duplex}
  \acro{fd}[FD]{full-duplex}
  \acro{sic}[SIC]{self-interference cancellation}
  \acro{si}[SI]{self-interference}
  \acro{bs}[BS]{base station}
  \acro{fbmc}[FBMC]{Filter Bank Multi-Carrier}
  \acro{ufmc}[UFMC]{Universal Filtered Multi-Carrier}
  \acro{scm}[SCM]{Single Carrier Modulation}
  \acro{isi}[ISI]{inter-symbol interference}
  \acro{ftn}[FTN]{Faster-Than-Nyquist}
  \acro{m2m}[M2M]{machine-to-machine}
  \acro{mtc}[MTC]{machine type communication}
  \acro{mmw}[mmWave]{millimeter wave}
  \acro{bf}[BF]{beamforming}
  \acro{los}[LOS]{line-of-sight}
  \acro{nlos}[NLOS]{non line-of-sight}
  \acro{capex}[CAPEX]{capital expenditure}
  \acro{opex}[OPEX]{operational expenditure}
  \acro{ict}[ICT]{information and communications technology}
  \acro{sp}[SP]{service providers}
  \acro{inp}[InP]{infrastructure providers}
  \acro{mvnp}[MVNP]{mobile virtual network provider}
  \acro{mvno}[MVNO]{mobile virtual network operator}
  \acro{nfv}[NFV]{network function virtualization}
  \acro{vnfs}[VNF]{virtual network functions}
  \acro{cran}[C-RAN]{Cloud Radio Access Network}
  \acro{bbu}[BBU]{baseband unit}
  \acro{bbus}[BBU]{baseband units}
  \acro{rrh}[RRH]{remote radio head}
  \acro{rrhs}[RRH]{Remote radio heads} 
  \acro{sfv}[SFV]{sensor function virtualization}
  \acro{wsn}[WSN]{wireless sensor networks} 
  \acro{bio}[BIO]{Bristol is open}
  \acro{vitro}[VITRO]{Virtualized dIstributed plaTfoRms of smart Objects}
  \acro{os}[OS]{operating system}
  \acro{www}[WWW]{world wide web}
  \acro{iotvn}[IoT-VN]{IoT virtual network}
  \acro{mems}[MEMS]{micro electro mechanical system}
  \acro{mec}[MEC]{Mobile edge computing}
  \acro{coap}[CoAP]{Constrained Application Protocol}
  \acro{vsn}[VSN]{Virtual sensor network}
  \acro{rest}[REST]{REpresentational State Transfer}
  \acro{aoi}[AoI]{Age of Information}
  \acro{lora}[LoRa\texttrademark]{Long Range}
  \acro{iot}[IoT]{Internet of Things}
  \acro{snr}[SNR]{Signal-to-Noise Ratio}
  \acro{cps}[CPS]{Cyber-Physical System}
  \acro{uav}[UAV]{Unmanned Aerial Vehicle}
  \acro{rfid}[RFID]{Radio-frequency identification}
  \acro{lpwan}[LPWAN]{Low-Power Wide-Area Network}
  \acro{lgfs}[LGFS]{Last Generated First Served}
  \acro{wsn}[WSN]{wireless sensor network} 
  \acro{lmmse}[LMMSE]{Linear Minimum Mean Square Error}
  \acro{rl}[RL]{Reinforcement Learning}
  \acro{nb-iot}[NB-IoT]{Narrowband IoT}
  \acro{lorawan}[LoRaWAN]{Long Range Wide Area Network}
  \acro{mdp}[MDP]{Markov Decision Process}
  \acro{ann}[ANN]{Artificial Neural Network}
  \acro{dqn}[DQN]{Deep Q-Network}
  \acro{mse}[MSE]{Mean Square Error}
  \acro{ml}[ML]{Machine Learning}
  \acro{cpu}[CPU]{Central Processing Unit}
  \acro{ddpg}[DDPG]{Deep Deterministic Policy Gradient}
  \acro{ai}[AI]{Artificial Intelligence}
  \acro{gp}[GP]{Gaussian Processes}
  \acro{drl}[DRL]{Deep Reinforcement Learning}
  \acro{mmse}[MMSE]{Minimum Mean Square Error}
  \acro{fnn}[FNN]{Feedforward Neural Network}
  \acro{eh}[EH]{Energy Harvesting}
  \acro{wpt}[WPT]{Wireless Power Transfer}
  \acro{dl}[DL]{Deep Learning}
  \acro{yolo}[YOLO]{You Only Look Once}
  \acro{mec}[MEC]{Mobile Edge Computing}
  \acro{marl}[MARL]{Multi-Agent Reinforcement Learning}

\acro{rl}[RL]{Reinforcement Learning}
\acro{drl}[DRL]{Deep Reinforcement Learning}
\acro{madrl}[MADRL]{Multi-Agent Deep Reinforcement Learning}
\acro{marl}[MARL]{Multi-Agent Reinforcement Learning}
\acro{dqn}[DQN]{Deep Q-Network}
\acro{ddpg}[DDPG]{Deep Deterministic Policy Gradient}
\acro{ppo}[PPO]{Proximal Policy Optimization}
\acro{mdp}[MDP]{Markov Decision Process}
\acro{ml}[ML]{Machine Learning}
\acro{cnn}[CNN]{Convolutional Neural Network}
\acro{lstm}[LSTM]{Long Short-Term Memory}
\acro{cpu}[CPU]{Central Processing Unit}
\acro{hpa}[HPA]{Horizontal Pod Autoscaling}
\acro{vpa}[VPA]{Vertical Pod Autoscaling}
\acro{laas}[LaaS]{Localization as a Service}
\acro{vm}[VM]{Virtual Machine}
\acro{qos}[QoS]{Quality of Service}
\acro{kpi}[KPI]{Key Performance Indicator}
\acro{sla}[SLA]{Service Level Agreement}

\acro{madqn}[MA-DQN]{Multi-Agent DQN}
\acro{maddpg}[MA-DDPG]{Multi-Agent DDPG}
\acro{mappo}[MA-PPO]{Multi-Agent PPO}
\acro{api}[API]{Application Programming Interface}
\acro{ble}[BLE]{Bluetooth Low Energy}
\acro{ai}[AI]{Artificial Intelligence}
\acro{vnf}[VNF]{Virtual Network Function}
\acro{nfv}[NFV]{Network Function Virtualization}


\acro{marise}[MARLISE]{Multi-Agent Reinforcement Learning-based In-place Scaling Engine}

\acro{mariseppo}[MARLISE]{Multi-Agent Reinforcement Intelligent Scaling Engine with PPO}
\acro{marisedqn}[MARLISE-DQN]{MARISE-DQN}
\acro{mariseddpg}[MARLISE-DDPG]{MARISE-DDPG}

\end{acronym}

\begin{abstract}
Modern edge-cloud systems face challenges in efficiently scaling resources to handle dynamic and unpredictable workloads. Traditional scaling approaches typically rely on static thresholds and predefined rules, which are often inadequate for optimizing resource utilization and maintaining performance in distributed and dynamic environments. This inefficiency hinders the adaptability and performance required in edge-cloud infrastructures, which can only be achieved through the newly proposed in-place scaling. To address this problem, we propose the Multi-Agent Reinforcement Learning-based In-place Scaling Engine (MARLISE) that enables seamless, dynamic, reactive control with in-place resource scaling. We develop our solution using two Deep Reinforcement Learning algorithms: Deep Q-Network (DQN),  and Proximal Policy Optimization (PPO). We analyze each version of the proposed MARLISE solution using dynamic workloads, demonstrating their ability to ensure low response times of microservices and scalability. Our results show that MARLISE-based approaches outperform heuristic method in managing resource elasticity while maintaining microservice response times and achieving higher resource efficiency.
\end{abstract}

\acresetall

\begin{IEEEkeywords}
Edge-cloud, Kubernetes, Auto-scaling, Multi-Agent Deep Reinforcement Learning, In-place scaling
\end{IEEEkeywords}

\copyrightnotice

\section{Introduction}
\label{sec:intro}


The cloud-native paradigm encompasses a set of principles and practices for designing, developing and managing software systems that fully leverage the capabilities of cloud computing~\cite{cn-survey}. It emphasizes key attributes such as scalability, performance and adaptability so that systems can dynamically adapt to varying loads. For example, a video streaming platform must scale its infrastructure to accommodate peak traffic during live events~\cite{madrl-bitrate-adaptation}, ensuring uninterrupted viewing experiences for users. Similarly, a cloud service running a \ac{ml} model might dynamically scale resources to make real-time predictions, such as identifying user location~\cite{10.1145/3581791.3596861} and activity patterns in a given area based on the time of day, enabling targeted services or recommendations. Furthermore, as cloud infrastructures are connected to the edge, resulting in so-called edge-cloud systems \cite{10.1145/3659097}, the infrastructure becomes more heterogeneous in terms of physical computing capability. 

These scenarios, involving dynamic user demands and heterogeneous, distributed infrastructure, highlight the need for adaptive and efficient resource management solutions. As modern microservices typically follow a microservice architecture, resource scaling can be done at different levels, such as the container, the pod, or the cluster~\cite{10.1145/3626246.3653378}.
For stateful microservices, i.e., microservices that maintain persistent data across requests, \ac{vpa} techniques that adjust the resource requests and limits (\ac{cpu} and memory) are more suitable compared to \ac{hpa}, which adjusts the number of pods based on demand. However, existing scaling methods primarily rely on predefined rules or reactive thresholds, which lack context awareness and slow to adapt to dynamic and unpredictable loads~\cite{10.1145/3603166.3632165}. Consequently, they struggle to handle fluctuations in request volume and state consistency requirements in stateful microservices. As a result, such solutions cannot optimize resource utilization in distributed environments, typically leading to under- or over-provisioning situations~\cite{10.1145/3626246.3653378}.
As also discussed by Coutinho \textit{et al.}~\cite{resource-elasticity-survey}, achieving resource elasticity is a fundamental challenge in environments where loads are highly dynamic and resources are limited.

To enable resource allocation for stateful microservices, in-place scaling was recently proposed as a method to adjust CPU and memory resources without requiring pod restarts, allowing services to scale more efficiently while maintaining microservice state~\cite{kubernetes-resize}. However, it has been noticed that there is a gap in the ability of existing \ac{vpa} tools to minimize resource slack and respond promptly to throttling of stateful microservices, leading to increased costs and impacting crucial metrics such as throughput and availability~\cite{10.1145/3626246.3653378}. Considering the dynamics of in-place scaling and the distributed, heterogeneous nature of edge-cloud \ac{vpa}, a \ac{madrl}-based solution~\cite{madrl-survey-apps, another-madrl-survey}  that relies on distributed agents that learn and adapt through experience, seems particularly suitable for the design of a scalable autoscaling engine. 
Decentralized \ac{madrl} approaches offer a promising solution by exploiting the modular structure of cloud-native systems. Each agent can control the resource allocation for a given microservice in a cloud-based environment, resulting in an intelligent and scalable solution. In addition, \ac{madrl} enables collaboration between agents to maintain system-wide performance while meeting the dynamic demands of workloads, making it an ideal candidate to address resource elasticity management in cloud-native environments for microservices in-place scaling.

In this paper we make the following contributions:

\begin{itemize}
\item We propose a novel \ac{vpa} solution for in-place scaling of stateful microservices, \ac{marise}, and develop it using two well-known \ac{drl} algorithms: \ac{dqn} and \ac{ppo} to develop both discrete and continuous versions of the solution.

\item  In our experimental evaluation, we demonstrate that \ac{marise} can stably and dynamically scale resources while effectively reducing \acp{kpi}, such as response time, compared to a conventional heuristic baseline when the load varies dynamically over time, i.e., when the number of requests per microservice fluctuates.

\item We show that the proposed solution can prioritize resources for specific microservices when necessary, whereas the heuristic approach is unable to do so. 

\item We also demonstrate that our solution is highly scalable and adapts seamlessly when stateful microservices are removed or added in the cloud.
\end{itemize}

The rest of the document is organized as follows: Section~\ref{sec:rw} provides a summary of related work, while Section~\ref{sec:background} outlines the background and key challenges. Section~\ref{sec:marlise} introduces and elaborates on \ac{marise}, the proposed solution. Section~\ref{sec:methodology} details the experimental evaluation methodology while Section~\ref{sec:eval} discusses the results. Finally, Section~\ref{sec:conclusions} concludes the paper.

\section{Related Work}
\label{sec:rw}

Efficient resource management remains a significant challenge in edge-cloud systems, especially given their heterogeneous infrastructure and highly dynamic workloads. Traditional scaling methods rely heavily on static rules or thresholds, which fail to swiftly adapt to fluctuating demands~\cite{resource-elasticity-survey}. To overcome these limitations, more advanced approaches utilizing \ac{ml}, such as \ac{cnn}\cite{sinan} and \ac{lstm}\cite{bilstmpaper, lstm_scaling}, have been proposed. Such approaches anticipate resource requirements based on historical data. However, these methods typically lack adaptability to sudden workload changes, which is critical in dynamic edge-cloud scenarios.


To that end, \ac{rl}-based approaches~\cite{hypredrl, Lotfi2023OpenRL, derp, eerlang, ResourceManagementDRL, prlov, 5g-resource-qlearning, dran, hpavpa} have emerged as promising solutions for real-time resource allocation. For example, the works in~\cite{hypredrl, Lotfi2023OpenRL} integrate \ac{lstm}-based predictions with \ac{rl} to enable more precise scaling decisions. The authors in~\cite{derp} proposed a Deep Elastic Resource Provisioning (DERP) approach and demonstrated improvements in \ac{vm} scaling using \ac{dqn}. Similarly, the Erlang autoscaler~\cite{eerlang} employs a multi-armed bandit algorithm to allocate \acp{vm} to microservices, while in~\cite{hpavpa}, \ac{rl} was employed to allocate resources(both horizontally  and vertically) in containerized cloud applications.
Unfortunately, none of these approaches consider \textbf{in-place scaling} required for stateful services, nor do they account for distributed decision-making. %
Other related efforts employed \ac{rl} for task scheduling~\cite{ResourceManagementDRL} and network management, such as slice admission control~\cite{prlov} or 5G radio access network slicing~\cite{5g-resource-qlearning}
and \ac{qos} optimization~\cite{dran}. 
To address distributed decision-making requirements, \ac{madrl} approaches have emerged to handle decentralized control challenges across various domains, including resource allocation in cloud computing environments~\cite{controller-scheduler-q-learning-hpa, mutliagentLinReg, marl_q_learning}, network management in 5G and beyond~\cite{drlec, distributed_marl_resource_allocation,  ma-multi-tentant}, and wireless networks~\cite{mutliagent-similar-to-ours}.
Nevertheless, existing \ac{madrl} methods have not yet been tailored to \textbf{in-place \ac{vpa}} for stateful microservices, thus failing to resolve critical limitations around uninterrupted and fine-grained resource allocation.

Table~\ref{tab:related_work} highlights how the proposed MARLISE method uniquely enables real-time, in-place scaling for stateful microservices compared to existing approaches. 
The proposed approach addresses identified gaps in the literature by employing \ac{madrl} specifically tailored for real-time in-place scaling, ensuring scalability, adaptability, and uninterrupted stateful microservice operations.


\begin{table}[t!]
    \centering
    \caption{Comparative Analysis of Scaling Approaches.}
    \begin{tabular}{lccccc}
        \toprule
        \makecell{\textbf{Approaches} \\ ---  \\  \textbf{Feature}} &  \makecell{\textbf{Traditional} \\ \textbf{ ML} \\ \cite{sinan, bilstmpaper, lstm_scaling}} & \makecell{\textbf{RL-based} \\ \cite{hypredrl, Lotfi2023OpenRL, derp, hpavpa, eerlang, 5g-resource-qlearning, ResourceManagementDRL, prlov, dran} } & \makecell{\textbf{MADRL} \\ \cite{controller-scheduler-q-learning-hpa, mutliagentLinReg, drlec, marl_q_learning, distributed_marl_resource_allocation, mutliagent-similar-to-ours, ma-multi-tentant} } & \makecell{ \textbf{Proposed} \\ \textbf{MARLISE} }\\
        \midrule
        \makecell{Real-time\\adaptation}  & \xmark & \cmark & \cmark & \cmark \\ \hline
        \makecell{Distributed \\ decision \\ making} & \xmark & \xmark & \cmark & \cmark \\ \hline
        \makecell{In-place \\  scaling} & \xmark & \xmark & \xmark & \cmark \\ \hline
        \makecell{Stateful \\  microservice \\ specific} & \xmark & \xmark & \xmark & \cmark \\ 
        
        \bottomrule
    \end{tabular}
    \label{tab:related_work}
    \vspace{-15pt}
\end{table}

\section{In-place Auto-scaling}
\label{sec:background}
In our work, we assume an edge-cloud environment realized through a cloud-native technology stack in which stateful applications realized as microservices are packaged as containers and managed by an orchestration platform. The most widely used and most versatile container orchestration solution is Kubernetes~\cite{kubernetes} and its flavours such as Microk8s~\cite{microk8s}.

\subsection{Resource Allocation in Kubernetes Clusters}

\begin{figure}[t]
    \centering
    \includegraphics[width=0.95\columnwidth]{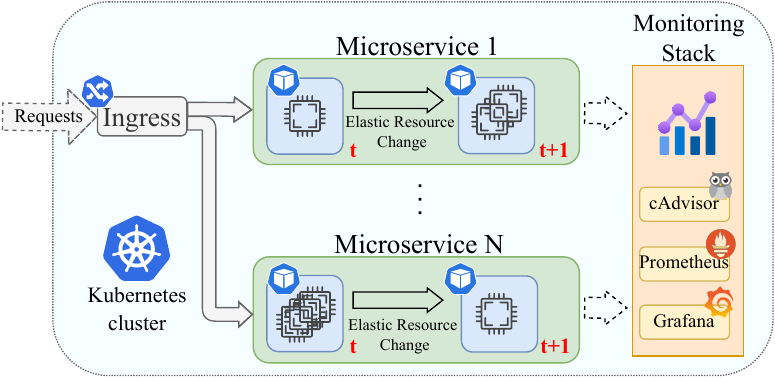}
    \caption{Kubernetes cluster hosting $N$ microservices running inside pods and dynamically adjusting resources over time. Between timestamps $t$ and $t+1$, microservice 1 increases its allocated resources, while microservice $N$ decreases its resources.
    }
    
    \label{fig:kubernetes-system-model}
    \vspace{-15pt}
\end{figure}

A Kubernetes cluster consists of a set of nodes that abstract the physical or virtual infrastructure and provide the environment for running containerized workloads. The cluster typically consists of worker nodes that host containerized applications and a master node (or control plane) that is responsible for managing the overall operation of the cluster, including orchestration and communication between the nodes. In this work, the term \textit{microservices} refers specifically to applications encapsulated in containers, i.e., application code with its dependencies running in Kubernetes pods. Each node provides a finite pool of resources, such as \ac{cpu} and memory, which is determined by the physical infrastructure of the host. These resources can be dynamically allocated to the microservices, as illustrated in Fig.~\ref{fig:kubernetes-system-model}.

In a Kubernetes environment, resources are shared between containers running on nodes and are initially managed by the Kubernetes scheduler, which allocates resources based on the specified requests and limits defined in the deployment configurations. Each container within a pod can request a certain amount of \ac{cpu} and memory to ensure it has the minimum necessary to function, while limits define the maximum it can consume. When resource contention occurs, Kubernetes uses \ac{qos} classes to prioritize containers and ensures that higher priority microservices get the resources they need, while the remaining resources are distributed to lower priority microservices. After deployment, traditional scaling methods adjust resource specifications either horizontally through \ac{hpa}, by adding or removing containers, or vertically through \textit{traditional \ac{vpa}}, by creating new containers with updated resources and removing the old containers. These approaches can lead to overhead, downtime, and inefficiencies. Nevertheless, the \ac{vpa} in Kubernetes can dynamically adjust the requests and limits values according to a pluggable algorithm \cite{10.1145/3626246.3653378}, so it is possible to create more intelligent and dynamic control logic. Furthermore, the newly proposed in-place resizing is able to dynamically adjust compute resources within a pod, without creating a new one and destroying the old one, a process that leads to interrupting microservices~\cite{kubernetes-resize}. These two enablers pave the way to improved \ac{vpa} techniques towards ensuring increased resource efficiency.

In order to ensure dynamic and intelligent \textit{in-place VPA} scaling, the control logic typically relies on various metrics that can be collected to monitor resource usage and application performance, such as \ac{cpu} and memory usage, disk I/O and network bandwidth. Tools such as Prometheus~\cite{prometheus} and cAdvisor~\cite{cadvisor} can be integrated into the cloud-native technology stack to collect such data from the edge-cloud system at regular intervals. These metrics can be used to make informed scaling decisions and can be visualized and analyzed with Grafana~\cite{grafana}.

\label{sec:allocation}

\subsection{Challenge of Dynamic Allocation of In-place Resources}
\label{sec:challenges}

\textit{In-place VPA} resource allocation in edge-cloud environments presents a difficult problem: \textit{How to allocate limited resources such as \ac{cpu} and memory to each microservice in a way that maintains acceptable performance across the system without interruptions?} To solve this problem we identify the following three challenges: 

\paragraph*{Adaptability to Dynamic Load Patterns} Effective \textit{in-place VPA} scaling requires complex real-time decisions to dynamically adjust resource allocations to ensure minimal response times within a few seconds and prevent performance degradation in microservices. Traditional static heuristics or rule-based approaches struggle to cope with the unpredictability of load patterns. To enable adaptive decision-making in real time, advanced online learning techniques, such as classification models or \ac{rl}, are essential. These methods enable the system to continuously learn from the evolving load and adapt resources autonomously.
	
\paragraph*{Priority management} Ensuring that microservices with high-priority receive sufficient resources while preventing microservices with lower-priority from experiencing resource depletion and becoming unresponsive is a non-trivial task. 
Resource contention in constrained environments requires intelligent arbitration, not only deciding how much to allocate, but also determining which services should take priority when resources are scarce. This requires a sophisticated context-aware allocation mechanism that takes into account (i) the global resource constraints of the system, (ii) the diverse and dynamic requirements of individual microservices, and (iii) the need to maintain system-wide response time guarantees.

\paragraph*{System scalability at maximum utilization} Microservices are elastic by nature, i.e., they can be dynamically instantiated or terminated as required. An \textit{in-place VPA} scaling solution must therefore support the reallocation of resources for newly introduced microservices, even if all available resources are already allocated. This becomes particularly difficult when a new microservice requires additional resources and the system is forced to reallocate resources from currently running microservices without causing a cascading performance degradation. Any effective solution must have self-adaptive capabilities to ensure that scaling decisions remain efficient and balanced across the entire system.



\paragraph*{Proposed MARLISE solution} To address these challenges in dynamic edge cloud environments, we adopt the \ac{madrl} framework, modeling each microservice as an independent learning agent that makes real-time scaling decisions. Unlike traditional heuristic-based or centralized methods, \ac{madrl} enables decentralized decision-making, reducing overhead and improving responsiveness. It continuously learns and adapts based on system feedback, effectively handling unpredictable loads. By integrating \ac{madrl} with in-place VPA scaling, we develop MARLISE, a self-optimizing, autonomous resource management system that enhances performance, scalability, and resource efficiency. The details of MARLISE are described in Section~\ref{sec:marlise}, while the next section examines the limitations of native auto-scaling solutions, providing the motivation for our proposed approach.

\subsection{Limitation of Native Auto-scaling Solutions}
\label{sec:limitaions}

Kubernetes offers two native approaches to auto-scaling resources: \ac{hpa}~\cite{kubernetes-hpa} and \ac{vpa}~\cite{kubernetes-vpa}. \ac{hpa} adjusts the number of pod replicas in a deployment based on observed CPU utilization or other selected metrics. \ac{vpa}, still in the experimental phase of Kubernetes autoscaler functionality, allocates resources (e.g., CPU or memory) to the pod based on historical utilization. While \ac{hpa} ensures that workloads can handle varying demand by scaling out or in, \ac{vpa} optimizes resource allocation per pod to improve efficiency and reduce wasted capacity. However, both methods require restarting the container, resulting in significant overhead and adding a time delay in response. In other words, the native solutions lack support for seamless scaling, a key feature our proposed solution has.

\begin{table}[t!]
    \centering
    \caption{Comparison of native auto-scaling methods (Heuristic) with the proposed MARLISE method.}
    \begin{tabular}{lcccc}
        \toprule
        \multirow{1}{*}{\textbf{Scaling Feature/Scaling Approach}} & \multicolumn{1}{c}{\textbf{\ac{hpa}}} & \multicolumn{1}{c}{\textbf{\ac{vpa}}} & \multicolumn{1}{c}{\textbf{\ac{marise}}} \\
        \midrule
        Time interval for scaling decision & $15s$ & $1m$ & $1s$ \\
        Best for \textbf{stateless} microservice & \cmark & \xmark & \xmark \\
        Best for \textbf{stateful} microservices & \xmark & \cmark & \cmark \\
        Support seamless scaling & \xmark & \xmark & \cmark \\
        Ability to resize pods & \xmark & \cmark & \cmark \\
        Adaptability to dynamic load patterns & \xmark & \xmark & \cmark \\
        Priority management & \xmark & \xmark & \cmark \\
        System scalability at maximum utilization & \xmark & \xmark & \cmark \\
        \bottomrule
    \end{tabular}
    \label{tab:scaling-details-kubernetes}
    \vspace{-15pt}
\end{table}

In Table~\ref{tab:scaling-details-kubernetes}, we outline the main differences between native solutions and the proposed \ac{marise} solution. Note that every approach supports both stateless and stateful microservices. However, \ac{marise} and \ac{vpa} perform better for stateful services, while \ac{hpa} is more effective for stateless services. Regarding the time interval for scaling decisions (i.e., the decision time-step), our proposed solution is able to make scaling decisions every second, whereas \ac{hpa} scales every 15 seconds, and \ac{vpa} scales pods approximately every minute. 
Furthermore, both \ac{marise} and \ac{vpa} can resize pods by adjusting allocated resources, whereas \ac{hpa} can only scale by adding additional pods, leading to coarser resource allocation. This makes \ac{vpa} and \ac{marise} more granular in resource adjustments.

Finally, native solutions are not designed to address the challenges of in-place resource allocation, as discussed in the previous subsection. In contrast, our proposed solution is specifically tailored to overcome these limitations. While proposed \ac{marise} takes advantage of newly proposed in-place resizing feature in Kubernetes, it introduces a learned, dynamic logic to optimize resource allocation, making it highly adaptable to microservice which load patterns change in a matter of seconds. Additionally, it ensures prioritization, efficient resource management, and system scalability.

\section{MADRL for In-place VPA Resource Allocation}
\label{sec:marlise}

\ac{rl} is, at its core, a sequential decision-making process in which agent learns to maximize cumulative rewards through interaction with the environment. This includes key concepts such as states, actions, and rewards, where the agent discovers a strategy (i.e., a set of actions) that maximizes the long-term reward. In the case of \ac{madrl}, multiple agents work together to improve coordination and dynamically allocate resources, prioritize resource allocation, and enable a scalable approach to managing the system, such as dynamically adding and removing resources such as allocating CPUs or memory for a microservice. Unlike traditional methods, \ac{madrl} uses trial-and-error interactions with the environment to learn resource allocation in real time without the need for centralized control.

\begin{figure}[t!]
	\centering
	\includegraphics[width=0.8\linewidth]{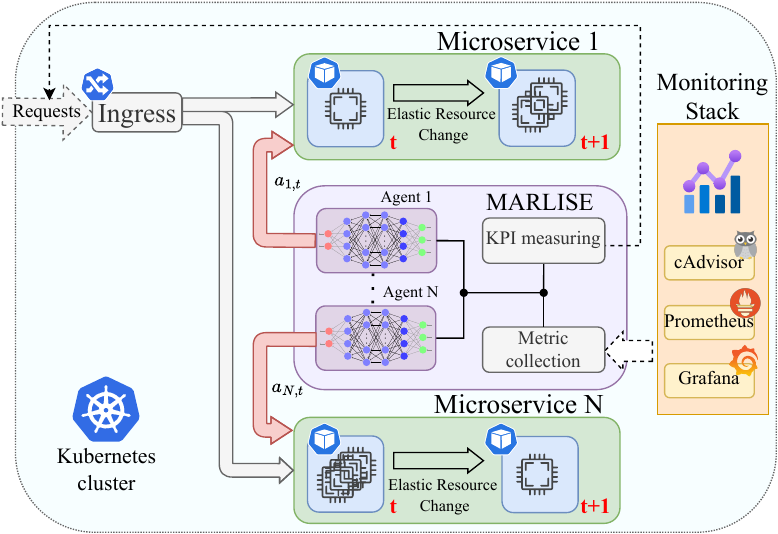}
   \caption{Illustration of the proposed MARLISE setup showing how agents manage resource allocation for each microservice.}

	\label{fig:agents-system-model}
 \vspace{-15pt}
\end{figure}

Dynamic workload management, a key challenge in such systems, requires real-time \textit{scaling decisions} to minimize response times and avoid microservice degradation. This is enabled by learning techniques such as \ac{drl}, which are well suited for adaptive and real-time decisions. Similarly, priority management ensures that high-demand microservices receive the necessary resources without neglecting others. This involves not only deciding how to allocate resources, but also which microservices should take precedence when resources are limited in order to balance the varying demands of each microservice while maintaining system-wide performance metrics such as response time. In addition, the scalability of the system requires the ability to dynamically add or remove microservices. \ac{madrl} addresses these challenges (see Section~\ref{sec:challenges}) by allowing agents to work independently and coordinate to effectively achieve system-level goals.

Each agent in the proposed framework monitors and allocates resources to a specific microservice within the system, as shown in Fig.~\ref{fig:agents-system-model}. The main goal of each agent is to dynamically in-place adjust the resource allocation for the assigned microservice, ensuring that \acp{kpi} such as response time and throughput are met while staying within predefined resource utilization thresholds. This decentralized approach increases scalability and flexibility as agents can manage their microservices independently while optimizing the system together.  
\ac{madrl} also enhances robustness, as the suboptimal behavior of one agent can be compensated by others, ensuring system-wide performance under varying conditions. 


We propose \textit{\ac{marise}-Discrete} and \textit{\ac{marise}-Continuous}, each tailored to the specific challenges of dynamic resource allocation, priority management and scalability in edge cloud environments, while enabling adaptive and efficient decision making across the system.


\subsection{States, Action, and Reward signal}
\label{sec:sar}

To ensure that the agents have sufficient information regarding the system, we design the \textbf{state} to encompass seven variables that describe the current snapshot of the system:
\begin{itemize}
 \item Resource Limit - The current allocation of resources;
 \item Resource Usage - Consumption, the gap between the limits and the consumption can indicate whether an increase or decrease is required;
 \item Available Resources - Availability, the agent is aware of the system's free resources;
 \item Utilization - Efficiency of the allocated resources;
 \item Average utilization of other resources - Knowledge about the resource utilization of other agents;
 \item Priority - Indicator of how important the microservice is;
 \item Average priority of other microservices - Helps the system to make trade-offs by taking into account the relative importance of all running services.
\end{itemize}

In addition, a time window of past variables is maintained, with each variable containing multiple past values stored in a First-In, First-Out (FIFO) order. This approach enables the model to leverage historical context for each variable, facilitating more informed decision-making based on recent system behavior. Specifically, the state space consists of $7 \times k$ input neurons, where $k$ represents the number of past values stored for each variable.
To enhance learning, state values are normalized, which enables relative resource allocation and consistency across different resource levels. This normalization improves the efficiency of \ac{drl} models by standardizing inputs and helping the model to focus on meaningful patterns and generalize optimal allocation strategies.

\textbf{Actions} are divided into two types: Discrete and continuous. \ac{marise}-Discrete operates in a discrete action space and selects from a set of predefined actions, such as increasing, decreasing or maintaining resource levels — ideal for straightforward, categorical adjustments. In contrast, \ac{marise}-Continuous  use a continuous action space and generate actions as float values in the range $[-1, 1]$. These values are scaled and applied as precise adjustments to resource allocations, allowing for granular and flexible control.

The \textbf{reward signal} 
balances two objectives: encouraging efficient \ac{cpu} utilization within desirable thresholds, and minimizing response times, particularly for high-priority microservices.
Let \( \eta_{i}(t) \) represent the utilization of the $i$-th microservice at time step \( t \), with upper and lower utilization thresholds denoted as $\eta_U$ and $\eta_L$ respectively.
The utilization reward $\rho_{i}(t)$ for each agent, which measures how efficiently the resources are being used at time step \( t \) and is defined as:
\begin{equation}\label{eq:reward-eta}
    \rho_{i}(t) = 
    \begin{cases}
    1 + \frac{\eta_{i}(t)}{\eta_U - \eta_L} & \text{if } \eta_L \leq \eta_{i}(t) \leq \eta_U \\
    \min\left(\frac{\Delta_{i}(t)}{10}, 1\right) & \text{if } \eta_{i}(t) < \eta_L \text{ and } \Delta_{i}(t) > 0 \\
    \max\left(\frac{\Delta_{i}(t)}{10}, -1\right) & \text{if } \eta_{i}(t) < \eta_L \text{ and } \Delta_{i}(t) \leq 0 \\
    \phantom{AAAA} 0 &   \phantom{AAAA}  \text{ Otherwise}
    \end{cases},
\end{equation}
where \( \Delta_{i}(t) = \eta_{i}(t) - \eta_{i}(t-1) \), 
represents the change in utilization compared to the previous timestep.
The Eq.~\ref{eq:reward-eta} discourages both low and excessive utilization while incentivizing higher utilization towards the upper threshold \( \eta_U \).

Let \( \omega(t) \) denote the weighted response time at time step \( t \), which contains the priority \( p_i \) and the response times \( \sigma_i \)
of each service:
\begin{equation}\label{eq:weighted-rt}
    \omega(t) = \sum_{i=1}^{N} (1 + p_i)\cdot \sigma_i(t).
\end{equation}

\noindent The weighted response time as defined in Eq. \ref{eq:weighted-rt} aggregates response times across all microservices, emphasizing those with higher priority. This ensures critical microservices have greater influence on the overall reward, promoting prioritization during resource allocation.


A \textbf{shared reward} \( r_s(t) \) is then defined to minimize the overall response times in the system and act as a centralized reward signal for collaboration between all agents:
\begin{equation}\label{eq:shared-reward}
    r_s(t) = 1 - \alpha \cdot (\omega(t) - 0.01), 
\end{equation}

\noindent with $\alpha$ denotes the scaling factor assigned to response time. A higher $\alpha$  places greater emphasis on overall system performance relative to individual agent performance.
According to above equation,
\( r_s(t) \) provides a positive reward when the response time approaches zero.
This shared reward encourages the agents to work together towards the system-wide goal of reducing response times.

The \textbf{reward} signal \( r(t) \) combines individual utilization efficiency with the overall response time performance, ensuring a balance between resource usage and microservice response time:
\begin{equation}\label{eq:final-reward}
    r_i(t) = \beta \cdot \rho_{i}(t) + r_s(t).
\end{equation}

\noindent The scaling factor $\beta \in(0, 1]$  captures how important agent utilization is in the reward signal. Note that the reward is determined after every agent in the system has selected an action and the system has transitioned to the next state.

\subsection{Multi-agent solutions}
\label{sec:mas}


1) \textit{\ac{marise} Discrete:} Is based on the \ac{dqn} approach~\cite{dqnPaper}. 
In the \ac{marise}-Discrete algorithm, each agent learns to approximate a Q-value function \( Q(s, a; \theta) \) using a deep neural network~\cite{dqnPaper}, where \( s \) represents the current state, \( a \) the selected action and \( \theta \) the parameters of the neural network.
Through repeated interactions with the environment, the agents learn the Q-values associated with each action-pair in such a way to be able to select the action with the highest expected reward in each state. For example, the agents add or remove a predetermined amount of CPU and memory to individual microservices. If the response time is high, resources are allocated to the microservice, while resources are removed if utilization is low.


As outlined in Alg.~\ref{alg:MultiAgentDQN}, each agent independently maintains a Q-network \( Q_i \), a target network \( \hat{Q}_i \), and an experience buffer \( \mathcal{D}_i \). 
The agents store transition \((s_i (t), a_i(t), r_i (t), s_i(t+1))\), i.e., experience, in the $\mathcal{D}_i$ buffer and then sample mini-batche of size $J$ to update their Q-networks when the experience buffer $\mathcal{D}_i$ reaches the defined size. The target network provides stable Q-value estimates using the Bellman equation (line~\ref{dqn:target-computation}). The Q-network is then trained by minimizing the loss between the current and target Q-values (line~\ref{dqn:update-polciy-network}), while a soft update of the target network (line~\ref{dqn:target-softupdate}) ensures stable learning and reduces the risk of policy divergence. 

\setlength{\textfloatsep}{0pt}
\begin{algorithm}[t]
\caption{Proposed \ac{marise}-Discrete algorithm}\label{alg:MultiAgentDQN}
\begin{algorithmic}[1]
\FOR{$i = 1$ to $N$ (agents)}
    \STATE Initialize experience buffer $\mathcal{D}_i$,  Q-network $Q_i(s, a; \theta_i)$ with random weights, and target Q-network $\hat{Q}_i(s, a; \theta_i^-)$ with weights $\theta_i^- = \theta_i$
\ENDFOR
\FOR{$k = 1$ to $K$ (episodes)}
    \FOR{$t = 1$ to $T$ (timesteps)}
        \FOR{$i = 1$ to $N$ (agents)}
            \STATE Select action $a_{i}(t)$ using $\epsilon$-greedy policy
            \STATE Execute action $a_{i}(t)$
            \STATE Observe the new state $s_{i}(t+1)$ and determine the reward $r_{i}(t)$ according to Eq.~\ref{eq:final-reward}
            \STATE Save transition $(s_{i}(t), a_{i}(t), r_{i}(t), s_{i}(t+1))$ 
            \STATE Randomly sample $J$ experiences from $\mathcal{D}_i$
            
            \FOR{every $\{ s_i(j), r(j), a_i(j), s_i(j+1) \}$ in batch}
        \STATE Set $y_i(j) = r(j) + $ \newline $\phantom{AAAAAAA}
        \gamma max_{a_i(j\text{+}1)} \hat{Q}_i(s_i(j\text{+}1), a_i(j\text{+}1))$ \label{dqn:target-computation}
        \ENDFOR
        \STATE  Calculate loss: \newline
        $\phantom{AAA} \mathcal{Z}_i = \frac{1}{J} \sum_{j=1}^{J} (Q_i(s_i(j), a_i(j)) - y_i(j) )^2$ \label{alg:proposed:line:15}
        \STATE Update $Q_i(s_i, a_i|\theta_i)$ by minimising the loss $\mathcal{Z}_i$ \label{dqn:update-polciy-network}
            \STATE Update target network:
            
            \hspace{2em}$\theta_i^- \leftarrow \tau \theta_i + (1 - \tau) \theta_i^-$ 
            \label{dqn:target-softupdate}
        \ENDFOR
    \ENDFOR
\ENDFOR
\end{algorithmic}
\end{algorithm}

2) \textit{\ac{marise} Continuous}: Is based on the \ac{ppo} approach~\cite{ppoPaper}. Each agent employs an actor-critic architecture, where the actor network selects actions and the critic network estimates state values to drive policy updates. The solution relies on a stochastic policy in continuous action spaces that allows agents to adaptively select actions based on probabilistic optimization. Each agent’s actor network \(\pi_i\) is individually updated by policy gradients to improve adaptability to changing states by performing gradient descent on Eq.~\ref{eq:clipped-surrogate-loss}. Note that resource allocation with \ac{marise}-Continuous adjustments are granular and occur within a predefined range. 

\ac{marise}-Continuous 
uses the vanilla advantage estimation method  (line~\ref{ppo:advantage-estimation} in Alg.~\ref{alg:MultiAgentPPO}) to compute the advantage function $\hat{A}_i$ for agent $i$, which measures how much better (or worse) performing a particular action \( a_i \) in a state \( s_i \) is compared to the average action in that state according to the current policy. 
The most important equation in the \ac{mappo} algorithm is the clipped surrogate loss in Eq.~\ref{eq:clipped-surrogate-loss}, which ensures that the new policy does not deviate significantly from the old one. This approach helps to maintain the stability of the training and avoids large, destructive updates. Let \( r_i(\theta) = \frac{\pi_{\theta_i}(a_i|s_i)}{\pi_{\theta_i^{\text{old}}}(a_i|s_i)} \) is the ratio of the probability that the agent $i$ takes the action \( a_i \) in state \( s_i \) in the new policy relative to the old policy. This ratio effectively measures the divergence between the policies. The \textbf{clipped surrogate loss} $\mathcal{L}_i^{\text{clip}}(\theta_i)$ for agent $i$ is then defined as:

\begin{equation}\label{eq:clipped-surrogate-loss}
    \mathbb{E} \left[ \min\left(r_i(\theta) \hat{A}_i, \text{clip}\left( r_i(\theta), 1 - \epsilon, 1 + \epsilon \right) \hat{A}_i \right) \right].
\end{equation}
The clipping mechanism restricts the policy update by limiting the ratio \( r_i(\theta) \) to the interval \([1 - \epsilon, 1 + \epsilon]\). This ensures that the updates do not push the new policy too far away from the old policy, reducing the risk of destabilizing the training process. Unlike the other proposed \ac{marise} methods, \ac{marise}-Continuous  discards past experience after each policy update and focuses solely on the most recent data to improve training relevance and responsiveness to current conditions in line~\ref{ppo:softupadte-delete-buffer}.

\setlength{\textfloatsep}{0pt}
\begin{algorithm}[t]
\caption{Proposed \ac{marise}-Continuous  algorithm}\label{alg:MultiAgentPPO}
\begin{algorithmic}[1]
\FOR{$i = 1$ to $N$ (agents)}
    \STATE Initialize actor network $\pi_{\theta_i}(a|s)$, critic network $V_{\phi_i}(s)$ with weights $\phi_i$ and buffer $\mathcal{D}_i$
\ENDFOR
\FOR{$m = 1$ to $M$ (episodes)}
    \FOR{$t = 1$ to $T$ (timesteps)}
        \FOR{$i = 1$ to $N$ (agents)}
            \STATE Sample action $a_{i}(t) \sim \pi_{\theta_i}(a|s_{i}(t))$
            \STATE Execute action $a_{i}(t)$
            \STATE Observe the new state $s_{i,{t+1}}$ and determine the reward $r_{i}(t)$ according to Eq.~\ref{eq:final-reward}
            \STATE Store transition $(s_{i}(t), a_{i}(t), r_{i}(t), s_{i}(t+1))$ in $\mathcal{D}_i$
        \ENDFOR
    \ENDFOR
    \FOR{$i = 1$ to $N$ (agents)}
        \STATE Compute $\hat{A}_i = R_i - V_{\phi_i}(s_i)$ for transitions in $\mathcal{D}_i$ \label{ppo:advantage-estimation}
        \STATE Compute $\mathcal{L}_i^{\text{clip}}(\theta_i)$ for transitions $\mathcal{D}_i$ as in Eq.~\ref{eq:clipped-surrogate-loss}

        \FOR{$k = 1$ to $K$ (epochs)} \label{ppo:k-epochs}
            \STATE Update $\theta_i$ by maximizing $\mathcal{L}_i^{\text{clip}}(\theta_i)$ 
            \label{ppo:actor-update}

            \STATE Update $\phi_i$ by minimizing the value loss:
            
            \hspace{2em}$L_V(\phi_i) = \mathbb{E}[(V_{\phi_i}(s_i) - R_i)^2]$
            \label{ppo:critic-loss}
        \ENDFOR
        \STATE Update $\theta_i^{\text{old}} \leftarrow \theta_i$ and clear buffer $\mathcal{D}_i$ \label{ppo:softupadte-delete-buffer}
    \ENDFOR
\ENDFOR
\end{algorithmic}
\end{algorithm}


\section{Evaluation Methodology}
\label{sec:methodology}

In this section, we describe the edge-cloud infrastructure, containerized microservice deployment, training process, and evaluation KPIs, followed by the heuristic baseline for comparison, which we use in the next evaluation section.

\subsection{Implementation and Realistic Deployment Considerations}

Dynamic in-place scaling is enabled by employing resource resizing feature of Kubernetes~\cite{kubernetes-resize}, allowing seamless \ac{cpu} and memory adjustments without container restarts. Communication with the Kubernetes cluster is implemented via the official Kubernetes Python client~\cite{kubernetes_python_client}, ensuring standard-compliant real-time resource adjustments. Real-time system metrics such as \ac{cpu} utilization, allocated resources, and available resources are collected every second using cAdvisor, integrated with Prometheus for data collection, and visualized using Grafana, accurately replicating realistic operational monitoring. Furthermore, the source code of the proposed \ac{marise} solution is publicly available for validation and replication purposes\footnote{The code supporting our experiments is publicly available on GitHub: https://github.com/sensorlab/agent-edge-autoscaling}.

Our experimental evaluation is conducted on an edge-cloud cluster comprising two Raspberry Pi 5 nodes (64-bit, 2.4 GHz quad-core ARM CPUs) acting as worker nodes, and one \ac{vm} equipped with a quad-core Xeon E5-2650 CPU at 2.00 GHz serving as the Kubernetes master node. Microk8s~\cite{microk8s} orchestrates the environment due to its lightweight and ARM compatibility, suitable for realistic edge-cloud deployments.




\subsection{Test Stateful Microservice and Workload Scenario}

The evaluation utilizes \textbf{\ac{laas}}, a representative \ac{ml}-driven microservice for device localization based on \ac{ble} beacons. \ac{laas} employs a fingerprinting method leveraging a pre-trained localization model to perform real-time inference, estimating device positions within a predefined grid. For realistic deployment, the \ac{laas} microservice is containerized using Docker~\cite{docker} and deployed within the Kubernetes environment, as demonstrated in our prior work~\cite{wcnc_demo}.

The microservice predominantly consumes \ac{cpu} resources, maintaining relatively constant memory utilization; thus, the evaluation explicitly focuses on \ac{cpu} resource allocation under dynamically varying workloads that emulate realistic \ac{iot} or smart-city scenarios, characterized by significant fluctuations in user demand patterns. More specifically, during the evaluation of dynamic load patterns (Section~\ref{sec:load_test}) and priority management (Section~\ref{sec:priority_test}), the system receives 100 requests per second, distributed as shown in Table~\ref{tab:load_table1}. Each request uses a \ac{laas} microservice to determine the location of the device. During the scalability evaluation (Section~\ref{sec:scalability_test}), Table~\ref{tab:load_table2} lists the number of requests received by each microservice.


\subsection{\ac{marise} Training}
The \textbf{training} of the \ac{marise} solutions was performed with a synthetic load at random intervals and with an intensity limited by an upper bound. This approach mimics the variability of real-world traffic (or load) while ensuring that the system operates within feasible resource limits so that agents can adapt effectively under dynamic conditions. Note that we use these pre-trained models in our evaluation.

\begin{table}[t!]
    \centering
    \caption{Microservice load levels during evaluation.}
    \begin{tabular}{lccc}
        \toprule
         \textbf{Time} ($s$) / \textbf{Microservice}  & \textbf{ 1} & \textbf{ 2} & \textbf{ 3} \\
        \midrule
        0--7  &  25.0\%   &  8.5\%    &  66.5\%     \\
        8--14 &  25.0\%    &  72.0\%   &  3.0\%    \\
        15--21 &  47.0\% &  7.0\% &  46.0\%     \\
        \bottomrule
    \end{tabular}
    \label{tab:load_table1}
\end{table}

\begin{table}[t!]
    \centering
    \caption{Request distribution for scalability evaluation.}
    \begin{tabular}{lcccc}
        \toprule
         \textbf{Time} ($s$) / \textbf{Microservice}  & \textbf{ 1} & \textbf{ 2} & \textbf{ 3}  & \textbf{4}\\
        \midrule
        0--15  &  30   &  40    &  \textemdash   &   \textemdash   \\
        16--30 &  30    &  40   &  50   &  \textemdash    \\
        31--45 &  30 &  40 &  50 &  50      \\
        46--60 &  \textemdash  &  40 &  50 &  50   \\
        \bottomrule
    \end{tabular}
    \label{tab:load_table2}
    \vspace{-5pt}
\end{table}

\begin{table}[t!]
    \centering
    \caption{\ac{marise} parameters.}
    \begin{tabular}{lccc}
        \toprule
        \multirow{1}{*}{\textbf{Parameter/Algorithm}} & \multicolumn{1}{c}{\textbf{Discrete}} & \multicolumn{1}{c}{\textbf{Continuous}} \\
        \midrule
Learning rate                     & $1 \times 10^{-4}$    & \textemdash           \\ 
Actor learning rate               & \textemdash           & $3 \times 10^{-4}$    \\ 
Critic learning rate              & \textemdash           & $1 \times 10^{-3}$    \\ 
Discount factor \( \gamma \)      & 0.99                  & 0.99                  \\ 
Replay buffer size                & 1000                  & \textemdash           \\ 
Batch $J$ size                        & 128                   & 600                   \\ 
Soft update parameter $\tau$      & 0.005                 & \textemdash           \\ 
Exploration tate                  & $\epsilon$ decay      & Variance              \\ 
Clip range                        & \textemdash           & 0.2                   \\ 
Entropy coefficient               & \textemdash           & 0.01                  \\ 
Update epochs ($K$)               & \textemdash           & 10                    \\ 
Num. of hidden layers             & 3                     & 4                     \\ 
Neurons per hid. layer            & $64-128-64$             & $64-128-128-64$         \\
Upper util. threshold $\eta_U$ & 60\% & 60\% \\
Lower util. threshold $\eta_L$ & 30\% & 30\% \\
Shared reward factor $\alpha$ & 5 & 5 \\
Util. reward factor  $\beta$ & 0.5 & 0.5 \\
        \bottomrule
    \end{tabular}
    \label{tab:all-algos-params}
\end{table}

Table~\ref{tab:all-algos-params} lists the fine-tuned parameters we use for the \ac{marise} solutions. Each exploration rate, i.e., $\epsilon$-decay, OU Noise and action variance, is set to decrease gradually. In this way, exploration is encouraged at first and after the first few training episodes, the system moves on to exploitation. The learning rates, the size of the replay buffers and the batch sizes have been optimized for stability and efficiency of performance.
The update epochs for \textit{\ac{marise}-Continuous} ($K=10$) were selected to maintain a balance between training stability and computational efficiency. 
he reward function scaling factors, $\alpha$ and $\beta$, defined in Eq.~\ref{eq:shared-reward} and Eq.~\ref{eq:final-reward}, along with the threshold parameters $\eta_L$ and $\eta_U$ (Eq.~\ref{eq:reward-eta}), were fine-tuned for the target microservice and deployment setting.

 
\subsection{KPIs}
\label{sec:kpi}
The \acp{kpi} evaluated included mean response time (the time it takes the cluster to process a request), violation rate (response times $>$ 250 ms), resource utilization 
(percentage of allocated resources used ) and resource deltas
, which represent the total resource changes. To obtain accurate and stable results for the evaluation experiments, the \ac{madrl} algorithms were executed and their results averaged over 20 iterations.


\subsection{Baseline for Comparison}


As we highlighted in Section~\ref{sec:limitaions}, native \ac{hpa} and \ac{vpa} are not designed for uninterrupted scaling. Therefore, a direct comparison with our method would be inherently unfair, as they would perform worse in every experiment. Furthermore, our proposed solution dynamically adapts to request fluctuations within seconds, a responsiveness that the native approaches lack. To ensure a fair evaluation, we compare our method to a \textbf{heuristic} approach that was developed based on the guidelines in~\cite{rzadca2020autopilot}. It uses a policy-driven technique that adjusts resource allocation based on CPU and memory usage trends and employs predefined thresholds to trigger scaling actions. This ensures dynamic resource adjustments while maintaining system stability. The method continuously monitors utilization to prevent under-allocation (which leads to performance degradation) and over-allocation (which is inefficient). Scaling decisions are made every second to ensure real-time responsiveness and a fair comparison with the proposed solution. These threshold-based scaling approaches, as used in Kubernetes' \ac{hpa} and \ac{vpa}, are computationally efficient but struggle with unpredictable load patterns in dynamic multi-service environments, as we show in the next section.

\section{Evaluation}
\label{sec:eval}

In this section, 
we evaluate the proposed \ac{marise} solution described in Section~\ref{sec:marlise}, according to the methodology in Section~\ref{sec:methodology}, demonstrating how it overcomes the challenges identified in Section~\ref{sec:challenges}.


\begin{table*}[htp]
    \centering
    \caption{Averaged performance (20 iterations) metrics for dynamic load experiment.}
    \begin{tabular}{lccccccccc}
        \toprule
        \multirow{2}{*}{\textbf{Metrics/Algorithm}} & \multicolumn{3}{c}{\textbf{Heuristic}} & \multicolumn{3}{c}{\textbf{\ac{marise}-Discrete}} & \multicolumn{3}{c}{\textbf{\ac{marise}-Continuous}} \\
        \cmidrule(lr){2-4} \cmidrule(lr){5-7} \cmidrule(lr){8-10} 
        \textbf{Microservices} & 1 & 2 & 3 & 1 & 2 & 3 & 1 & 2 & 3 \\
        \midrule
        Violations (\%) & \red{10.43} & 19.60 & \red{25.99} & \green{\textbf{7.14}} & \red{21.12} & 24.53 & 8.67 & \green{\textbf{19.26}} & \green{\textbf{12.05}} \\
        Mean response time ($s$) & \red{0.09} & \green{\textbf{0.22}} & 0.28 & \green{\textbf{0.08}} & \red{0.24} & \red{0.31} & \green{\textbf{0.08}} & \green{\textbf{0.22}} & \green{\textbf{0.14}}  \\
        Mean resource delta (mc) & \green{\textbf{415}} & \green{\textbf{488}} & \green{\textbf{578}} & \red{645} & 615 & 648 & 532 & \red{709} & \red{667}  \\
        \bottomrule
    \end{tabular}
    \label{tab:performance_metrics_loadtesting}
    \vspace{-5pt}
\end{table*}

\begin{table*}[h!]
    \centering
    \caption{Averaged performance (20 iterations) metrics for different priority experiment.}
    
    \begin{subtable}[t]{\textwidth}
        \caption{\textbf{Low High Medium.}}
        \centering
        \begin{tabular}{lcccccccccccc}
            \toprule
            \multirow{2}{*}{\textbf{Metrics/Algorithm}} & \multicolumn{3}{c}{\textbf{Heuristic}} & \multicolumn{3}{c}{\textbf{\ac{marise}-Discrete}} & \multicolumn{3}{c}{\textbf{\ac{marise}-Continuous}} \\
            \cmidrule(lr){2-4} \cmidrule(lr){5-7} \cmidrule(lr){8-10} \cmidrule(lr){11-13}
            \textbf{Microservices} & 1 (L) & 2 (H) & 3 (M) & 1 (L) & 2 (H) & 3 (M) & 1 (L) & 2 (H) & 3 (M) \\
            \midrule
            Violations (\%) & 10.43 & 19.60 & \red{25.99} & \green{\textbf{8.45}} & \red{19.63} & 24.63 & \red{12.18} & \green{\textbf{17.89}} & \green{\textbf{21.06}} \\
            Mean response time ($s$) & \green{\textbf{0.09}} & 0.22 & 0.28 & \green{\textbf{0.09}} & \green{\textbf{0.21}} & \red{0.30} & \red{0.14} & \red{0.29} & \green{\textbf{0.27}}  \\
            Mean resource delta (mc) & \red{415} & 488 & 578 & \green{\textbf{264}} & \green{\textbf{475}} & \red{598} & 359 & \red{574} & \green{\textbf{557}}  \\
            
            \bottomrule
        \end{tabular}
    \end{subtable}

    \begin{subtable}[t]{\textwidth}
    \vspace{2pt}
        \caption{\textbf{Medium Low High.}}
        \centering
        \begin{tabular}{lcccccccccccc}
            \toprule
            \multirow{2}{*}{\textbf{Metrics/Algorithm}} & \multicolumn{3}{c}{\textbf{Heuristic}} & \multicolumn{3}{c}{\textbf{\ac{marise}-Discrete}} & \multicolumn{3}{c}{\textbf{\ac{marise}-Continuous}} \\
            \cmidrule(lr){2-4} \cmidrule(lr){5-7} \cmidrule(lr){8-10} \cmidrule(lr){11-13}
            \textbf{Microservices} & 1 (M) & 2 (L) & 3 (H) & 1 (M) & 2 (L) & 3 (H) & 1 (M) & 2 (L) & 3 (H) \\
            \midrule
            Violations (\%) & \red{10.43} & 19.60 & \red{25.99} & \green{\textbf{7.47}} & \green{\textbf{19.56}} & 23.46 & 9.53 & \red{20.87} & \green{\textbf{16.81}} \\
            Mean response time ($s$) & 0.09 & \green{\textbf{0.22}} & \red{0.28} & \green{\textbf{0.08}} & 0.24 & 0.26 & \red{0.11} & \red{0.34} & \green{\textbf{0.21}}  \\
            Mean resource delta (mc) & \red{415} & {488} & 578 & \green{\textbf{356}} & \green{\textbf{366}} & \red{664} & 374 & \red{499} & \green{\textbf{542}}  \\
            \bottomrule
        \end{tabular}
    \end{subtable}

    \begin{subtable}[t]{\textwidth}
    \vspace{2pt}
        \caption{\textbf{High Medium Low.}}
        \centering
        \begin{tabular}{lcccccccccccc}
            \toprule
            \multirow{2}{*}{\textbf{Metrics/Algorithm}} & \multicolumn{3}{c}{\textbf{Heuristic}} & \multicolumn{3}{c}{\textbf{\ac{marise}-Discrete}} & \multicolumn{3}{c}{\textbf{\ac{marise}-Continuous}} \\
            \cmidrule(lr){2-4} \cmidrule(lr){5-7} \cmidrule(lr){8-10} \cmidrule(lr){11-13}
            \textbf{Microservices} & 1 (H) & 2 (M) & 3 (L) & 1 (H) & 2 (M) & 3 (L) & 1 (H) & 2 (M) & 3 (L) \\
            \midrule
            Violations (\%) & \red{10.43} & \green{\textbf{19.60}} & 25.99 & \green{\textbf{7.10}} & 20.11 & \green{\textbf{18.66}} & 8.17 & \red{20.29} & \red{30.50} \\
            Mean response time ($s$) & \red{0.09} & \green{\textbf{0.22}} & 0.28 & \green{\textbf{0.08}} & \green{\textbf{0.22}} & \green{\textbf{0.26}} & \red{0.09} & \red{0.30} & \red{0.38}  \\
            Mean resource delta (mc) & 415 & 488 & \red{578} & \red{504} & \green{\textbf{399}} & \green{\textbf{444}} & \green{\textbf{413}} & \red{571} & 550  \\
            \bottomrule
        \end{tabular}
    \end{subtable}
    \label{tab:perfomrance_metrics_priorities_updated}
    \vspace{-6pt}
\end{table*}

\begin{table*}[h!]
    \centering
    \caption{Averaged performance (20 iterations) metrics for the scalability experiment.}
    \begin{subtable}[t]{\textwidth}
    \vspace{-3pt}
        \caption{\textbf{Agent 3 added.}}
        \centering
        \begin{tabular}{lcccccccccccccccc}
            \toprule
            \multirow{2}{*}{\textbf{Metrics/Algo.}} & \multicolumn{4}{c}{\textbf{Heuristic}} & \multicolumn{4}{c}{\textbf{\ac{marise}-Discrete}} & \multicolumn{4}{c}{\textbf{\ac{marise}-Continuous}} \\
            \cmidrule(lr){2-5} \cmidrule(lr){6-9} \cmidrule(lr){10-13} \cmidrule(lr){14-17}
            \textbf{Microservices} & 1 & 2 & 3 & 4 & 1 & 2 & 3 & 4 & 1 & 2 & 3 & 4\\
            \midrule
            Violations (\%) & 0.00 & 0.00 & \red{23.40} & \textemdash & 0.00 & 0.00 & \green{\textbf{17.03}} & \textemdash & 0.00 & 0.00 & 22.21 & \textemdash \\
            Mean response time ($s$) & 0.02 & 0.02 & 0.27 & \textemdash & 0.02 & 0.02 & \green{\textbf{0.26}} & \textemdash & 0.02 & 0.02 & \red{0.37} & \textemdash \\
            Mean resource delta (mc) & \green{\textbf{29}} & \green{\textbf{44}} & 305 & \textemdash & \red{450} & \red{475} & \red{370} & \textemdash & 102 & 96 & \green{\textbf{237}} & \textemdash \\
            \bottomrule
        \end{tabular}
    \end{subtable}
    
    \begin{subtable}[t]{\textwidth}
    \vspace{2pt}
        \caption{\textbf{Agent 4 added.}}
        \centering
        \begin{tabular}{lcccccccccccccccc}
            \toprule
            \multirow{2}{*}{\textbf{Metrics/Algo.}} & \multicolumn{4}{c}{\textbf{Heuristic}} & \multicolumn{4}{c}{\textbf{\ac{marise}-Discrete}} & \multicolumn{4}{c}{\textbf{\ac{marise}-Continuous}} \\
            \cmidrule(lr){2-5} \cmidrule(lr){6-9} \cmidrule(lr){10-13} \cmidrule(lr){14-17}
            \textbf{Microservices} & 1 & 2 & 3 & 4 & 1 & 2 & 3 & 4 & 1 & 2 & 3 & 4 \\
            \midrule
            Violations (\%) & 0.00 & 0.00 & 0.32 & 81.90 & 0.00 & 0.00 & \green{\textbf{0.00}} & \green{\textbf{45.99}} & 0.00 & 0.00 & 0.00 & 81.86 \\
            Mean response time ($s$) & 0.02 & 0.02 & 0.02 & 1.69 & 0.02 & 0.02 & 0.02 & \green{\textbf{0.71}} & 0.02 & 0.02 & 0.02 & 1.42  \\
            Mean resource delta (mc) & \green{\textbf{2}} & \green{\textbf{1}} & \green{\textbf{6}} & 7 & 131 & \red{244} & \red{262} & \red{139} & \red{147} & 171 & 176 & 66  \\
            \bottomrule
        \end{tabular}
    \end{subtable}
    
    \begin{subtable}[t]{\textwidth}
    \vspace{2pt}
        \caption{\textbf{Agent 1 removed.}}
        \centering
        \begin{tabular}{lcccccccccccccccc}
            \toprule
            \multirow{2}{*}{\textbf{Metrics/Algo.}} & \multicolumn{4}{c}{\textbf{Heuristic}} & \multicolumn{4}{c}{\textbf{\ac{marise}-Discrete}} & \multicolumn{4}{c}{\textbf{\ac{marise}-Continuous}} \\
            \cmidrule(lr){2-5} \cmidrule(lr){6-9} \cmidrule(lr){10-13} \cmidrule(lr){14-17}
            \textbf{Microservices} & 1 & 2 & 3 & 4 & 1 & 2 & 3 & 4 & 1 & 2 & 3 & 4 \\
            \midrule
            Violations (\%) & \textemdash & 0.00 & 0.00 & \red{18.67} & \textemdash & 0.00 & 0.00 & \green{\textbf{0.00}} & \textemdash & 0.00 & 0.00 & 11.11\\
            Mean response time ($s$) & \textemdash & 0.02 & 0.02 & \red{0.21} & \textemdash & 0.02 & 0.02 & \green{\textbf{0.03}} & \textemdash & 0.02 & 0.02 & 0.09 \\
            Mean resource delta (mc) & \textemdash & \green{\textbf{5}} & 42 & \red{195} & \textemdash & 99 & \red{143} & \green{\textbf{115}} & \textemdash & 35 & \green{\textbf{38}} & 170 \\
            \bottomrule
        \end{tabular}
    \end{subtable}

    \label{tab:performance_metrics_scalability_scenarios}
    \vspace{-15pt}
\end{table*}


\subsection{Evaluating Adaptability to Dynamic Load Patterns}\label{sec:load_test}

\begin{figure}[t!]
    \centering
    \includegraphics[width=0.99\linewidth]{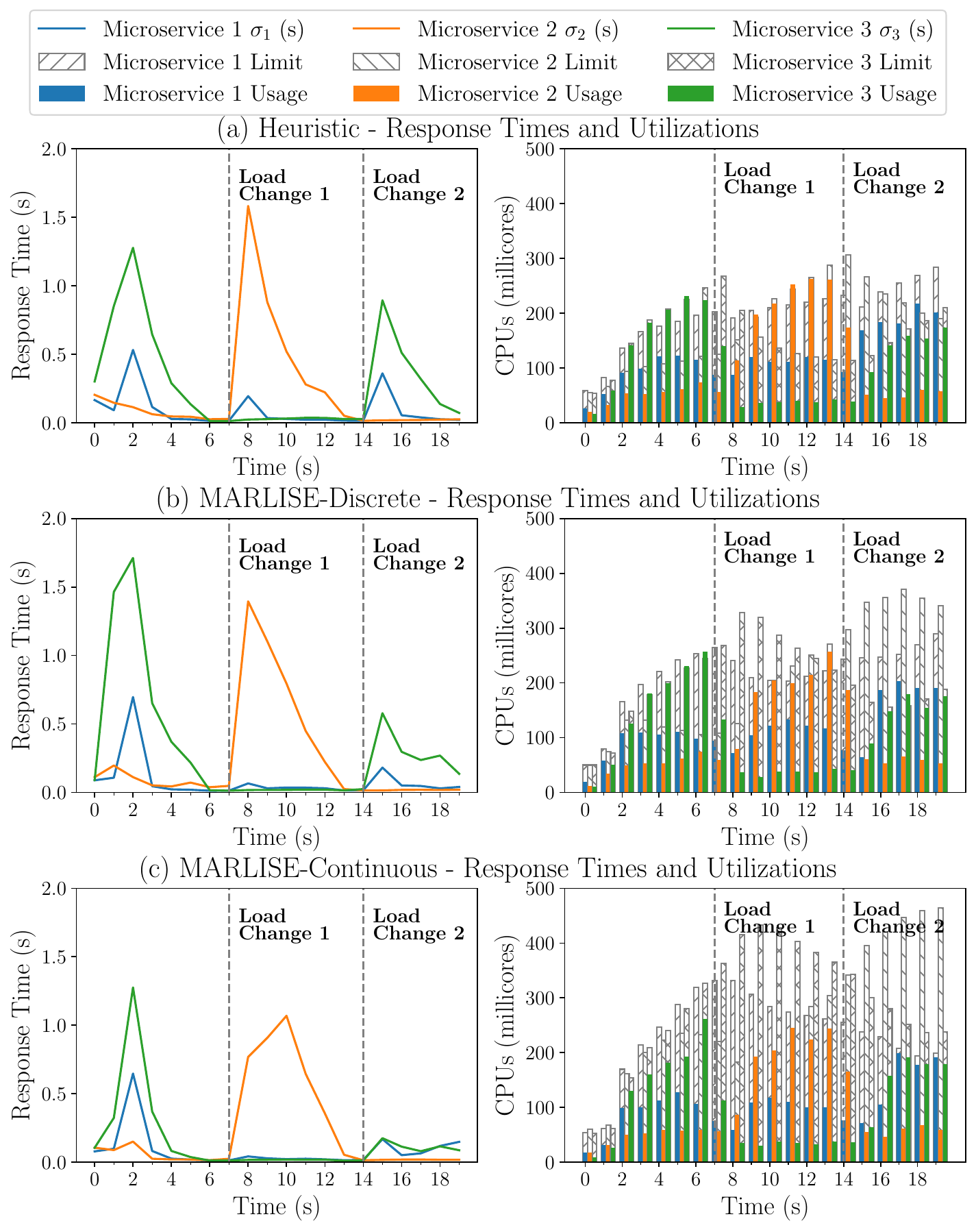}
    \vspace{-15pt}
    \caption{Averaged performance of different algorithms under dynamic load changes.}
    \label{fig:loadtesting_alg}
\end{figure}

In the first experiment, we evaluate how effectively the proposed solutions and the heuristic baseline respond to dynamic changes in load, i.e., addressing the first challenge of in-place scaling described in Section~\ref{sec:challenges}.
To collect performance data, the system is subjected to a workload of parallel requests per second, simulating user interactions with three \acp{laas}.
Fig.~\ref{fig:loadtesting_alg} shows the performance of the heuristic approach and the \ac{marise} algorithms under two load changes between microservices, as described in Table~\ref{tab:load_table1}.
The figures present the response time $\sigma_i$ of each service  $i$ and \ac{cpu} utilization over time and illustrate the effectiveness of each algorithm in allocating resources for the respective \ac{laas}, which mainly relies on \ac{cpu} resources for operation.

The results shown in Fig.~\ref{fig:loadtesting_alg} illustrate how the different algorithms react to dynamic changes in the number of requests to individual services. As can be seen from the response time graphs, each solution is able to adapt to changes within a few seconds. Among them, \ac{marise}-Continuous has the fastest response time compared to \ac{marise}-Discrete and the heuristic approach. 
A more detailed performance evaluation can be found in Table~\ref{tab:performance_metrics_loadtesting}, which contains the averaged results over 20 experimental iterations. These results include the \acp{kpi} described in Section~\ref{sec:kpi} and show significant performance differences between the algorithms. For example, in terms of violations and mean response time for microservice 3, the proposed Continuous solution shows superior performance, achieving a 0.14$s$ response time and the lowest violations at 12.05\%. 
In contrast, Discrete and heuristic achieve much higher response times (0.31s and 0.28s) and violations (24.53\% and 25.99\%), respectively. However, when considering the total amount of resources allocated to microservice 3 throughout the experiment, the baseline heuristic allocates the least resources (578mc), while Discrete allocates 648mc and Continuous allocates 667mc. This illustrates a fundamental trade-off between speed and resource efficiency: both \ac{marise}-Discrete and even more \ac{marise}-Continuous prioritize fast response times, but do so at the cost of more frequent resource allocation. This indicates that the heuristic approach is more conservative in resource allocation, resulting in higher response times and more violations.

\subsection{Evaluating Priority Management}\label{sec:priority_test}



In our second experiment, we focus on determining how well the proposed solutions can adapt to priority management, which corresponds to the second challenge outlined in Section~\ref{sec:challenges}. To evaluate the priority management, the system was subjected to the microservice load distribution described in Table~\ref{tab:load_table1}, with priority values $p_i$ assigned to the agents depending on the importance of each microservice. The reward signal defined in Eq.~\ref{eq:weighted-rt} encourages resource allocation according with these priorities.


Table~\ref{tab:perfomrance_metrics_priorities_updated} displays performance metrics for the \acp{kpi} listed in Section \ref{sec:kpi} in different combinations of priorities of their microservices. In Table~\ref{tab:perfomrance_metrics_priorities_updated}~(a), the priorities set in the \ac{marise} algorithms ensure that Microservice~2, which has the highest priority, consistently receives the highest share of \ac{cpu} resources to maintain low response times even when the load changes. Microservice~1, which has the lowest priority, receives only minimal resources, which leads to slower response times under heavy load. However, \ac{marise}-Discrete allocates resources effectively and achieves the best response times and the fewest violations. In addition, Table~\ref{tab:perfomrance_metrics_priorities_updated}~(b) illustrates similar performance when a different microservice priority is high. The results also show that the heuristic method has the highest percentage of violations, while the \ac{marise} solutions optimize the system-wide metrics even when complying to the priority settings. 

Table~\ref{tab:perfomrance_metrics_priorities_updated} shows that \ac{marise}-Discrete and \ac{marise}-Continuous best maintain the priority settings and effectively balance the \acp{kpi}. Our proposed solutions also allocate resources according to the set importance of the microservice, thus successfully balancing response times and resource utilization. Additionally, the results indicate that the heuristic method only serves as a useful baseline, as it does not respond to priority changes. These results suggest that \ac{marise}-Discrete and \ac{marise}-Continuous are much better suited for environments that require precise prioritization and adaptive scaling, and highlight the strengths of employing \ac{drl} in managing complex, priority-driven workloads.

\subsection{Evaluating system scalability at maximum utilization}\label{sec:scalability_test}


\begin{figure}[t!]
    \centering
    \includegraphics[width=0.99\linewidth]{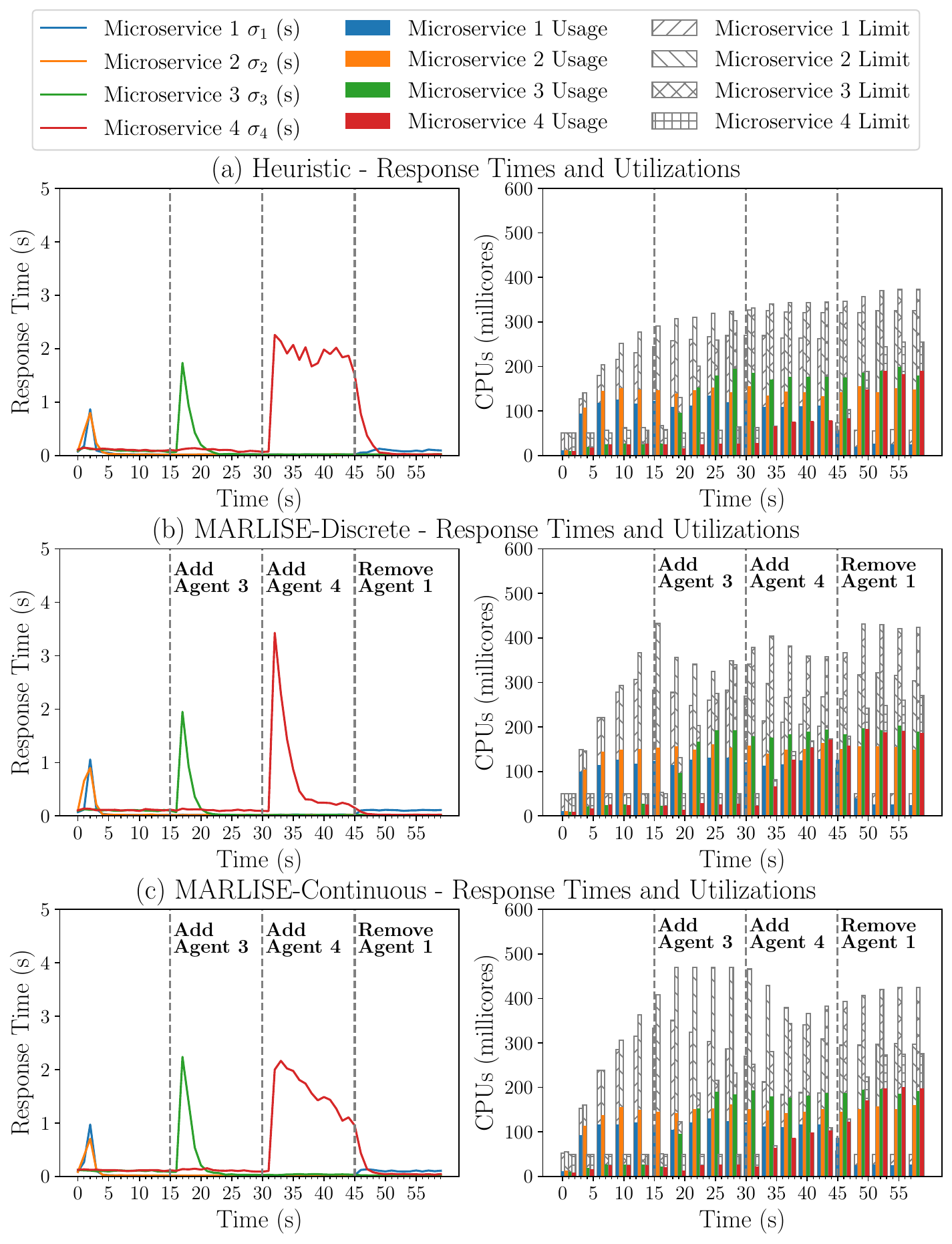}
    \caption{Scalability experiment with dynamic addition and removal of agents.}
    \label{fig:scalability_algs}
\end{figure}

In our last experiment, we evaluate the scalability, i.e., the ability to add and remove agents, of the proposed in-place scaling solution. The experiment starts with two agents, adds a third agent after 15 seconds and introduces a fourth agent 15 seconds later, following the microservice load distribution outlined in Table~\ref{tab:load_table2}. Finally, the first agent is removed to simulate a Kubernetes environment where microservices change frequently. This experiment tests the ability of each algorithm to manage limited resources under dynamic conditions and evaluates their responsiveness to changes in demand and agent configuration, even when resources are already fully allocated. Fig.~\ref{fig:scalability_algs} shows how each algorithm responds to these changes in real time.

At the start of the experimenent, every approach handles the workload well. However, the heuristic approach, as shown in Fig.~\ref{fig:scalability_algs}(a), has significant problems when agent~4, i.e. the fourth microservice, is added, as the response times become extremely high, indicating the limited scalability of the heuristic approach when the existing resources are already allocated and additional microservices are added. In contrast, \ac{marise} solution, as shown in Fig.~\ref{fig:scalability_algs}(b) and Fig.~\ref{fig:scalability_algs}(c) respectively, demonstrate significant better performance. Proposed solutions efficiently allocate resources to the additional agents and stabilize quickly after the initial peaks. This is particularly noticeable in the response times for Microservice~4, while the \ac{cpu} utilization for all microservices remains balanced. 

Such a behavior is also confirmed in Table~\ref{tab:performance_metrics_scalability_scenarios}, which shows the average performance over 20 iterations for metrics for every solution. The results also show that \ac{marise}-Discrete achieves the best results for violations, response times and resource deltas. These deltas indicate efficient collaboration between agents as resources are dynamically added or removed to optimize allocation without over provisioning. In each interval, \ac{marise}-Discrete stands out for its responsiveness to scaling changes, closely followed by \ac{marise}-Continuous, both of which exhibit adaptive and stable performance.

The scalability tests show that \ac{marise} is an effective algorithm for managing dynamic edge-cloud environments with microservice changes under limited resources. \ac{marise}-Discrete is characterized by maintaining low response times through fast adaptation, which makes it ideal for stable loads, while the adaptability of \ac{marise}-Continuous is beneficial in fluctuating conditions. The heuristic approach that served as the baseline, lacks the adaptability required for informed scaling decisions. These results highlight the potential of \ac{marise} solution for scalable, adaptive resource management in edge cloud systems.

\section{Conclusion}
\label{sec:conclusions}


In this paper, we proposed a \ac{madrl}-based in-place VPA scaling engine and developed two scaling solutions based on \ac{drl} algorithms: \ac{marise}-Discrete and \ac{marise}-Continuous. We evaluated the proposed solutions in three scenarios: Adaptability to dynamic load changes, priority management, and scaling the system when system resources are fully utilized. Our results show the effectiveness of the proposed solutions in managing dynamic workloads. Moreover, our results have shown that \ac{marise}-Continuous ensures stable resource allocation by avoiding overutilization. For priority allocations, both \ac{marise}-Discrete and \ac{marise}-Continuous effectively satisfied the \acp{kpi}, with \ac{marise}-Discrete exhibiting the best scalability, closely followed by \ac{marise}-Continuous. Overall, the \ac{marise} solution outperformed the heuristic method by retaining the set \acp{kpi} and improving system performance. Future work will focus on extending \ac{drl} to manage additional edge cloud scenarios, considering \ac{ml} scheduling techniques, and refining \ac{marise} for seamless integration and universal deployment.






\section*{Acknowledgements}


This work was funded in part by the Slovenian Research Agency (grants P2-0016 and MN-0009-106), by grant PID2023-148104OB-C43 funded by MICIU/AEI/10.13039/501100011033 and co-funded by ERDF/EU, by the European Commission NANCY project (No. 101096456), and by the HORIZON-MSCA-PF project TimeSmart (No. 101063721).
\balance

\bibliographystyle{IEEEtran}
\bibliography{IEEEabrv,bibliography}

\begin{thebibliography}{10}
\providecommand{\url}[1]{#1}
\csname url@samestyle\endcsname
\providecommand{\newblock}{\relax}
\providecommand{\bibinfo}[2]{#2}
\providecommand{\BIBentrySTDinterwordspacing}{\spaceskip=0pt\relax}
\providecommand{\BIBentryALTinterwordstretchfactor}{4}
\providecommand{\BIBentryALTinterwordspacing}{\spaceskip=\fontdimen2\font plus
\BIBentryALTinterwordstretchfactor\fontdimen3\font minus \fontdimen4\font\relax}
\providecommand{\BIBforeignlanguage}[2]{{%
\expandafter\ifx\csname l@#1\endcsname\relax
\typeout{** WARNING: IEEEtran.bst: No hyphenation pattern has been}%
\typeout{** loaded for the language `#1'. Using the pattern for}%
\typeout{** the default language instead.}%
\else
\language=\csname l@#1\endcsname
\fi
#2}}
\providecommand{\BIBdecl}{\relax}
\BIBdecl

\bibitem{cn-survey}
N.~Kratzke and P.~C. Quint, ``{Understanding cloud-native applications after 10 years of cloud computing - A systematic mapping study},'' \emph{Journal of Systems and Software}, vol. 126, pp. 1--16, 2017.

\bibitem{madrl-bitrate-adaptation}
H.~Wang, Z.~Long, H.~Dong, and A.~El~Saddik, ``{MADRL-Based Rate Adaptation for 360° Video Streaming With Multiviewpoint Prediction},'' \emph{IEEE Internet Things J.}, vol.~11, no.~15, pp. 26\,503--26\,517, 2024.

\bibitem{10.1145/3581791.3596861}
N.~Garg and N.~Roy, ``{Sirius: A Self-Localization System for Resource-Constrained IoT Sensors},'' in \emph{Proc. MobiSys 2023}.\hskip 1em plus 0.5em minus 0.4em\relax New York, NY, USA: Association for Computing Machinery, 2023, p. 289–302.

\bibitem{10.1145/3659097}
P.~Souza, T.~Ferreto, and R.~Calheiros, ``{Maintenance Operations on Cloud, Edge, and IoT Environments: Taxonomy, Survey, and Research Challenges},'' \emph{ACM Comput. Surv.}, vol.~56, no.~10, Jun. 2024.

\bibitem{10.1145/3626246.3653378}
A.~Pavlenko, J.~Cahoon, Y.~Zhu, B.~Kroth, M.~Nelson, A.~Carter, D.~Liao, T.~Wright, J.~Camacho-Rodr\'{\i}guez, and K.~Saur, ``{Vertically Autoscaling Monolithic Applications with CaaSPER: Scalable Container-as-a-Service Performance Enhanced Resizing Algorithm for the Cloud},'' in \emph{Proc. SIGMOD/PODS 2024}.\hskip 1em plus 0.5em minus 0.4em\relax New York, NY, USA: Association for Computing Machinery, 2024, p. 241–254.

\bibitem{10.1145/3603166.3632165}
A.~Rubak and J.~Taheri, ``{Machine Learning for Predictive Resource Scaling of Microservices on Kubernetes Platforms},'' in \emph{Proc. IEEE/ACM 16th International Conference on Utility and Cloud Computing}, ser. UCC '23.\hskip 1em plus 0.5em minus 0.4em\relax New York, NY, USA: Association for Computing Machinery, 2024.

\bibitem{resource-elasticity-survey}
E.~F. Coutinho, F.~R. de~Carvalho~Sousa, P.~A.~L. Rego, D.~G. Gomes, and J.~N. de~Souza, ``{Elasticity in cloud computing: a survey},'' \emph{Annals of Telecommunications}, vol.~70, no.~7, pp. 289--309, 2015.

\bibitem{kubernetes-resize}
\BIBentryALTinterwordspacing
Kubernetes, ``{Kubernetes Documentation: Resize a Container's Resources},'' 2023, accessed: 2024-11-28. [Online]. Available: \url{https://kubernetes.io/docs/tasks/configure-pod-container/resize-container-resources}
\BIBentrySTDinterwordspacing

\bibitem{madrl-survey-apps}
W.~Du and S.~Ding, ``{A survey on multi-agent deep reinforcement learning: from the perspective of challenges and applications},'' \emph{Artificial Intelligence Review}, vol.~54, no.~5, pp. 3215--3238, 2021.

\bibitem{another-madrl-survey}
T.~T. Nguyen, N.~D. Nguyen, and S.~Nahavandi, ``{Deep Reinforcement Learning for Multiagent Systems: A Review of Challenges, Solutions, and Applications},'' \emph{IEEE Trans. Cybern.}, vol.~50, no.~9, pp. 3826--3839, 2020.

\bibitem{sinan}
Y.~Zhang, W.~Hua, Z.~Zhou, G.~E. Suh, and C.~Delimitrou, ``{Sinan: ML-based and QoS-aware resource management for cloud microservices},'' in \emph{Proc. 26th ACM Int. Conf. ASPLOS}.\hskip 1em plus 0.5em minus 0.4em\relax New York, NY, USA: Association for Computing Machinery, 2021, pp. 167--181.

\bibitem{bilstmpaper}
N.-M. Dang-Quang and M.~Yoo, ``{Deep Learning-Based Autoscaling Using Bidirectional Long Short-Term Memory for Kubernetes},'' \emph{Applied Sciences}, vol.~11, no.~9, 2021.

\bibitem{lstm_scaling}
M.~P. Yadav, {Rohit}, and D.~K. Yadav, ``{Resource Provisioning Through Machine Learning in Cloud Services},'' \emph{Arabian Journal for Science and Engineering}, vol.~47, no.~2, pp. 1483--1505, 2022.

\bibitem{hypredrl}
P.~Liu, W.~Zhao, B.~Zhang, and J.~Wang, ``{Hybrid Elastic Scaling Strategy for Container Cloud based on Load Prediction and Reinforcement Learning},'' \emph{Journal of Physics: Conference Series}, vol. 2732, p. 012014, 2024.

\bibitem{Lotfi2023OpenRL}
F.~Lotfi and F.~Afghah, ``{Open RAN LSTM Traffic Prediction and Slice Management Using Deep Reinforcement Learning},'' \emph{57th Asilomar Conference on Signals, Systems, and Computers}, pp. 646--650, 2023.

\bibitem{derp}
C.~Bitsakos, I.~Konstantinou, and N.~Koziris, ``{DERP: A Deep Reinforcement Learning Cloud System for Elastic Resource Provisioning},'' in \emph{Proc. IEEE Int. Conf. Cloud Comput. Technol. Sci.}, 2018, pp. 21--29.

\bibitem{eerlang}
V.~Sachidananda and A.~Sivaraman, ``{Erlang: Application-Aware Autoscaling for Cloud Microservices},'' in \emph{Proc. 19th European Conference on Computer Systems}.\hskip 1em plus 0.5em minus 0.4em\relax New York, NY, USA: Association for Computing Machinery, 2024, pp. 888--923.

\bibitem{ResourceManagementDRL}
H.~Mao, M.~Alizadeh, I.~Menache, and S.~Kandula, ``{Resource Management with Deep Reinforcement Learning},'' in \emph{Proc. 15th ACM Workshop on Hot Topics in Networks}, 2016, pp. 50--56.

\bibitem{prlov}
S.~Saxena and K.~M. Sivalingam, ``{DRL-Based Slice Admission Using Overbooking in 5{G} Networks},'' \emph{IEEE Open Journal of the Communications Society}, vol.~4, pp. 29--45, 2023.

\bibitem{5g-resource-qlearning}
Y.~Shi, Y.~E. Sagduyu, and T.~Erpek, ``{Reinforcement Learning for Dynamic Resource Optimization in 5{G} Radio Access Network Slicing},'' in \emph{Proc. IEEE 25th Int. Workshop CAMAD}, 2020, pp. 1--6.

\bibitem{dran}
S.~K. Kasi, U.~S. Hashmi, S.~Ekin, A.~Abu-Dayya, and A.~Imran, ``{D-RAN: A DRL-Based Demand-Driven Elastic User-Centric RAN Optimization for 6G \& Beyond},'' \emph{IEEE Trans. Cogn. Commun. Netw.}, vol.~9, no.~1, pp. 130--145, 2023.

\bibitem{hpavpa}
F.~Rossi, M.~Nardelli, and V.~Cardellini, ``{Horizontal and Vertical Scaling of Container-Based Applications Using Reinforcement Learning},'' in \emph{Proc. IEEE 12th CLOUD}, 2019, pp. 329--338.

\bibitem{controller-scheduler-q-learning-hpa}
S.~M.~R. Nouri, H.~Li, S.~Venugopal, W.~Guo, M.~He, and W.~Tian, ``{Autonomic decentralized elasticity based on a reinforcement learning controller for cloud applications},'' \emph{Future Generation Computer Systems}, vol.~94, pp. 765--780, 2019.

\bibitem{mutliagentLinReg}
C.~G. Ralha, A.~H. Mendes, L.~A. Laranjeira, A.~P. Araújo, and A.~C. Melo, ``{Multiagent system for dynamic resource provisioning in cloud computing platforms},'' \emph{Future Generation Computer Systems}, vol.~94, pp. 80--96, 2019.

\bibitem{marl_q_learning}
A.~Belgacem, S.~Mahmoudi, and M.~Kihl, ``{Intelligent multi-agent reinforcement learning model for resources allocation in cloud computing},'' \emph{J. King Saud Univ. Comput. Inf. Sci.}, vol.~34, no. 6, Part A, pp. 2391--2404, 2022.

\bibitem{drlec}
J.~Chen, J.~Chen, and H.~Zhang, ``{DRLEC: Multi-agent DRL based Elasticity Control for VNF Migration in SDN/NFV Networks},'' in \emph{Proc. 26th IEEE Asia-Pacific Conf. Commun.}, 2021, pp. 89--93.

\bibitem{distributed_marl_resource_allocation}
J.~Menard, A.~Al-Habashna, G.~Wainer, and G.~Boudreau, ``{Distributed Resource Allocation In 5{G} Networks With Multi-Agent Reinforcement Learning},'' in \emph{Proc. ANNSIM}.\hskip 1em plus 0.5em minus 0.4em\relax IEEE, 2022, pp. 802--813.

\bibitem{ma-multi-tentant}
I.~Vilà, J.~Pérez-Romero, O.~Sallent, and A.~Umbert, ``{A Multi-Agent Reinforcement Learning Approach for Capacity Sharing in Multi-Tenant Scenarios},'' \emph{IEEE Trans. Veh. Technol.}, vol.~70, no.~9, pp. 9450--9465, 2021.

\bibitem{mutliagent-similar-to-ours}
N.~Naderializadeh, J.~Sydir, M.~Simsek, and H.~Nikopour, ``{Resource Management in Wireless Networks via Multi-Agent Deep Reinforcement Learning},'' in \emph{Proc. IEEE 21st Int. Workshop Sig. Proc. Adv. Wir. Com.}, 2020, pp. 1--5.

\bibitem{kubernetes}
\BIBentryALTinterwordspacing
Google, ``Kubernetes,'' 2014. [Online]. Available: \url{https://kubernetes.io/}
\BIBentrySTDinterwordspacing

\bibitem{microk8s}
\BIBentryALTinterwordspacing
C.~Ltd., ``{MicroK8s - Lightweight Kubernetes},'' 2014. [Online]. Available: \url{https://microk8s.io/}
\BIBentrySTDinterwordspacing

\bibitem{prometheus}
\BIBentryALTinterwordspacing
O.~Soruce, ``Prometheus,'' 2012. [Online]. Available: \url{https://prometheus.io/}
\BIBentrySTDinterwordspacing

\bibitem{cadvisor}
\BIBentryALTinterwordspacing
Google, ``{cAdvisor - Container Advisor},'' 2014. [Online]. Available: \url{https://github.com/google/cadvisor}
\BIBentrySTDinterwordspacing

\bibitem{grafana}
\BIBentryALTinterwordspacing
G.~Labs, ``{Grafana - Open-Source Observability Platform},'' 2014. [Online]. Available: \url{https://grafana.com/}
\BIBentrySTDinterwordspacing

\bibitem{kubernetes-hpa}
\BIBentryALTinterwordspacing
Kubernetes, ``{Horizontal Pod Autoscaler},'' Documentation, Technical Report, 2015. [Online]. Available: \url{https://kubernetes.io/docs/tasks/run-application/horizontal-pod-autoscale/}
\BIBentrySTDinterwordspacing

\bibitem{kubernetes-vpa}
\BIBentryALTinterwordspacing
------, ``{Vertical Pod Autoscaler},'' Documentation, GitHub, Technical Report, 2017. [Online]. Available: \url{https://github.com/kubernetes/autoscaler/tree/master/vertical-pod-autoscaler}
\BIBentrySTDinterwordspacing

\bibitem{dqnPaper}
\BIBentryALTinterwordspacing
V.~Mnih, K.~Kavukcuoglu, D.~Silver, A.~Graves, I.~Antonoglou, D.~Wierstra, and M.~Riedmiller, ``{Playing Atari with Deep Reinforcement Learning},'' 2013. [Online]. Available: \url{https://arxiv.org/abs/1312.5602}
\BIBentrySTDinterwordspacing

\bibitem{ppoPaper}
\BIBentryALTinterwordspacing
J.~Schulman, F.~Wolski, P.~Dhariwal, A.~Radford, and O.~Klimov, ``{Proximal Policy Optimization Algorithms},'' 2017. [Online]. Available: \url{https://arxiv.org/abs/1707.06347}
\BIBentrySTDinterwordspacing

\bibitem{kubernetes_python_client}
\BIBentryALTinterwordspacing
O.~Source, ``Kubernetes python client,'' GitHub repository. [Online]. Available: \url{https://github.com/kubernetes-client/python}
\BIBentrySTDinterwordspacing

\bibitem{docker}
\BIBentryALTinterwordspacing
I.~Docker, ``Docker,'' 2013. [Online]. Available: \url{https://www.docker.com/}
\BIBentrySTDinterwordspacing

\bibitem{wcnc_demo}
J.~Prodanov, B.~Bertalanič, C.~Fortuna, and J.~Hribar, ``{Demonstrating Smart Scaling of AI-Services for Future Networks},'' in \emph{Proc. IEEE WCNC 2025}, 2025, pp. 1--3.

\bibitem{rzadca2020autopilot}
K.~Rzadca, P.~Findeisen, J.~Swiderski, P.~Zych, P.~Broniek, J.~Kusmierek, P.~Nowak, B.~Strack, P.~Witusowski, S.~Hand \emph{et~al.}, ``{Autopilot: Workload Autoscaling at Google Scale },'' in \emph{Proc. EuroSys 2020}, 2020, pp. 1--16.

\end{thebibliography}

\end{document}